\documentclass[conference]{IEEEtran}
\ifCLASSINFOpdf
\else
\fi
\usepackage{graphicx}
\usepackage{setspace}
\usepackage{tablefootnote}
\usepackage{amsmath,empheq}
\usepackage{amssymb}
\usepackage{amsthm}
\usepackage{caption}
\usepackage{subcaption}
\usepackage{bm} 
\usepackage{cite}
\usepackage{float}
\usepackage[usenames,dvipsnames]{pstricks}
\usepackage{epsfig}
\usepackage{pst-grad} 
\usepackage{pst-plot} 
\usepackage{tabularx}
\usepackage[]{flushend}
\usepackage{amsmath}
\usepackage[T1]{fontenc}

\newcommand{\nn} {\nonumber}
\usepackage{xcolor}
\usepackage{lipsum}
\usepackage{comment}

\allowdisplaybreaks

\allowdisplaybreaks

\usepackage{caption}
\usepackage[utf8]{inputenc}

\begin{document}


\title{Location Privacy for ISAC Systems: Rate, Sensing, and Privacy Tradeoff}

\author{
\IEEEauthorblockN{
Umair~Ali~Khan\IEEEauthorrefmark{1},
Lester~Ho\IEEEauthorrefmark{1},
Holger~Claussen\IEEEauthorrefmark{1}\IEEEauthorrefmark{2}\IEEEauthorrefmark{3},
Mark~F.~Flanagan\IEEEauthorrefmark{4},
and~Chinmoy~Kundu\IEEEauthorrefmark{1}
}

\IEEEauthorblockA{\IEEEauthorrefmark{1}Wireless Communications Laboratory, Tyndall National Institute, Dublin, Ireland}
\IEEEauthorblockA{\IEEEauthorrefmark{2}School of Computer Science and Information Technology, University College Cork, Cork, Ireland}
\IEEEauthorblockA{\IEEEauthorrefmark{3}Electronic \& Electrical Engineering, Trinity College Dublin, Dublin, Ireland}
\IEEEauthorblockA{\IEEEauthorrefmark{4}School of Electrical and Electronic Engineering, University College Dublin, Belfield, Ireland}

\IEEEauthorblockA{\{umairali.khan, lester.ho, holger.claussen, chinmoy.kundu\}@tyndall.ie, and mark.flanagan@ieee.org}
}

\maketitle
\pagestyle{plain} 

\begin{abstract} In this paper, we design a beamformer for transmitter location privacy in an integrated sensing and communication (ISAC) system, where the transmitter communicates with a receiver while sensing a point target. We adopt the direction-of-arrival obfuscation power ratio (DoA-OPR), the ratio of the received power in a prescribed fake direction to that in the actual direction, as the location privacy metric. To design the beamformer, we formulate an achievable rate maximization problem subject to constraints on total transmit power, target sensing, location privacy, and the number of data streams. The resulting problem is nonconvex and is addressed through an eigenmode-selection and power-allocation framework. To reduce the computational complexity of the solution, we develop strategies that eliminate infeasible candidate solutions. 
Furthermore, we propose a fake angle selection technique in which the transmit power for target sensing aids location privacy. 
Numerical results illustrate the achievable rate--sensing--privacy tradeoff and show that the proposed fake angle selection technique performs better than the random fake angle selection technique.
\end{abstract}

\section{Introduction}
\label{sec:introduction}

Integrated sensing and communication (ISAC) has emerged as a promising technology for sixth-generation wireless networks \cite{liu2022isac}. By sharing spectrum, hardware, and signal processing resources, ISAC jointly supports data transmission and environmental sensing, improving resource utilization and reducing deployment costs. 
However, sensing capabilities also introduce location-privacy
risks, multi-antenna receivers can exploit received
signals to localize transmitters \cite{checa2020location}.
This motivates ISAC designs that protect against unauthorized localization while maintaining communication and sensing performance.






Physical-layer location privacy has been investigated in multiple-input multiple-output (MIMO) communication systems to protect transmitters against unauthorized localization \cite{checa2020location, privacy_our_paper, zhang2025privacy, li2024fakepath}. In \cite{checa2020location}, transmit beamforming  conceals the angle-of-departure of each path while suppressing the line-of-sight (LoS) path to achieve transmitter location privacy. In \cite{privacy_our_paper},  a new privacy metric called the direction-of-arrival obfuscation power ratio (DoA-OPR), the ratio of the received angular power in a prescribed fake direction to that in the actual direction, is introduced to design a transmit beamformer that does not suppress the LoS path for location privacy. In \cite{zhang2025privacy}, a transmit beamformer is designed for a MIMO orthogonal frequency division multiplexing system to minimize the Cram\'er--Rao bound (CRB) of legitimate localization while the CRB of illegitimate localization is kept above a specific threshold. 
In \cite{li2024fakepath}, fake paths are  injected to impair DoA-based localization by an unauthorized receiver.

Although it is important, location privacy has received limited attention in ISAC scenarios \cite{ma2024sensing, chen2026scatterers, jia2025illegal}. In \cite{ma2024sensing},  a beamformer is designed for a MIMO transmitter to conceal its DoA from an ISAC receiver.
However, the system does not include simultaneous target sensing by the location privacy enabled transmitter. Furthermore, the reflected echo signal from any node in an ISAC system can lead to unauthorized location sensing of that node \cite{chen2026scatterers, jia2025illegal}. 
To protect the target’s location privacy, the ISAC transmitter in \cite{chen2026scatterers} intentionally illuminates known scatterers to deceive eavesdroppers into misidentifying scatterers as targets. 
In \cite{jia2025illegal}, an ISAC transmitter serves communication users and senses an adversary. To protect the user's location privacy, beamforming and artificial noise are jointly designed to impair the adversary's ability to estimate the user's direction using reflected signals.

The above articles \cite{chen2026scatterers} and \cite{jia2025illegal} consider the location privacy of the target or receiver in ISAC scenarios;  however, they do not consider the location privacy of the ISAC transmitter itself. 
To the best of our knowledge, transmit beamformer design that achieves location privacy of an ISAC transmitter while simultaneously sensing a target has not been investigated. 
Motivated by this, we consider a MIMO ISAC transmitter that communicates with a semi-trusted receiver while sensing a point target. The semi-trusted receiver is authorized to decode the transmitted data, but may also attempt to infer the transmitter's location without authorization.
The main contributions of this work are listed as follows.
\begin{itemize}
    \item We design a transmit beamformer for an ISAC transmitter in a single-user, single-target scenario to maximize the achievable rate subject to target sensing and transmitter location privacy constraints.


\item  To characterize the CRB–rate region for each privacy threshold, we determine two corner points on each boundary by minimizing the sensing CRB and maximizing the communication rate, respectively. By varying the privacy threshold, we show a three-way tradeoff between rate, sensing, and location privacy.

    \item We develop strategies that eliminate infeasible candidate solutions
    before optimization to reduce computational complexity.

    \item We propose a fake-angle selection technique in which transmit power for target sensing aids location privacy, thereby improving the achievable rate.
    
\end{itemize}


\textit{Notation:}
$(\cdot)^T$ and $(\cdot)^H$ denote transpose and conjugate
transpose, respectively. $\bar z$ denotes the conjugate of $z$. $\mathbb E[\cdot]$, $\operatorname{tr}(\cdot)$,
and $\det(\cdot)$ denote expectation, trace, and determinant.
$\mathbf I_N$ is the $N\times N$ identity matrix, and
$\mathbf X\succeq\mathbf0$ indicates that $\mathbf X$ is
Hermitian positive semidefinite.
$\mathcal{CN}(\boldsymbol\mu,\boldsymbol\Sigma)$ denotes a complex Gaussian distribution with mean
$\boldsymbol\mu$ and covariance $\boldsymbol\Sigma$.



\section{System Model}
\label{sec:system}
\begin{figure}
\centering
\includegraphics[width=0.4\linewidth]{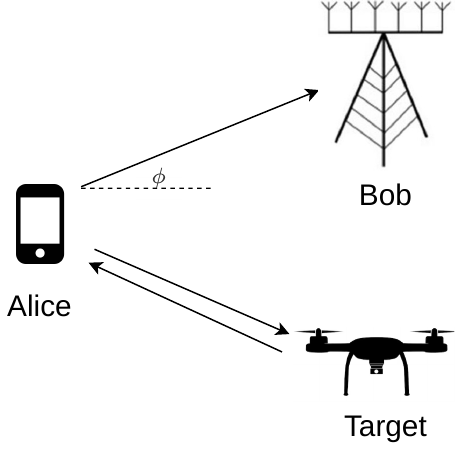}
\caption{System Model}
\label{P2systemmodel}
\vspace{-.5cm}
\end{figure}

We consider a narrowband MIMO ISAC system where Alice with $N_\mathrm{T}$ transmit antennas communicates with Bob equipped with $N_\mathrm{R}$ receive antennas,  as shown in Fig. \ref{P2systemmodel}. Alice also employs $N_\mathrm{S}$ receive antennas for simultaneously sensing a far-field point target. The communication channel between Alice and Bob and the  sensing channel between Alice and the target are denoted as $\mathbf H_\mathrm{c}\in\mathbb C^{N_\mathrm{R}\times N_\mathrm{T}}$ and $\mathbf H_\mathrm{s}\in\mathbb C^{N_\mathrm{S}\times N_\mathrm{T}}$, respectively. Next, we discuss the communication, sensing, and privacy models considered in the paper. 


\subsection{Communication Model}

\label{sec_comm_model}
We assume that both Alice and Bob employ antennas in the form of a
uniform linear array (ULA). The normalized array steering
vector for the transmit ULA can be written as \cite{alkhateeb},
\begin{align}
\label{eq_steering_vector}
    \mathbf{a}_\textrm{T}(\phi) = \frac{1}{\sqrt{N_\textrm{T}}} \left[1, e^{-j\frac{2\pi d}{\lambda} \sin(\phi)}, \ldots, e^{-j\frac{2\pi d}{\lambda}(N_\textrm{T}-1)\sin(\phi)}\right]^T,
\end{align}
where $\phi\in [-90,90]^\circ$ is the direction of departure
(DoD), $d$ is the spacing between antennas, and $\lambda$ denotes the wavelength of operation. The array steering vector of the receiver, $\mathbf{a}_\textrm{R}(\phi)$, is defined analogously.



Alice transforms the power-normalized complex baseband signal $\mathbf s(n)   \sim\mathcal{CN}(\mathbf0,\mathbf I_{N_\textrm{d}})$ with $N_\textrm{d}\le\min(N_\mathrm{T},N_\mathrm{R})$ independent data streams using a  precoder $\mathbf{W} \in \mathbb{C}^{N_\textrm{T} \times N_\textrm{d}}$ into the complex transmit signal $\mathbf{x}(n)=\mathbf{W} \mathbf{s}(n) \in \mathbb{C}^{N_\textrm{T}\times1}$ at each symbol index $n$. The precoding matrix $\mathbf{W}$ satisfies the transmit power constraint $\text{tr}(\mathbf{W}\mathbf{W}^H) = P$. The signal $\mathbf s(n)$ employs independent symbols in each symbol index $n$. The transmit covariance of $\mathbf{x}(n)$ is
\begin{equation}
\mathbf Q=\mathbb E[\mathbf x(n)\mathbf x^H(n)]=\mathbf W\mathbf W^H,
\quad \mathbf Q\succeq\mathbf0,
\label{eq:covariance}
\end{equation}
where  $\operatorname{rank}( \mathbf {Q}) =N_\textrm{d}$.
The received signal at Bob can be written as
\begin{align}
\mathbf y_\textrm{c}(n)=\mathbf H_\textrm{c}\mathbf x(n)+\mathbf z_\textrm{c}(n),
\label{eq:comm_signal}
\end{align}
where $\mathbf z_\textrm{c}(n)$ denotes the independent additive white Gaussian noise vector at Bob, with distribution $\mathcal{CN}(\mathbf0,\sigma_\textrm{c}^2\mathbf I_{N_\mathrm{R}})$.

We adopt the Rician channel model for the communication channel between Alice and Bob, i.e.,
\begin{equation}
\mathbf H_\textrm{c}=\sqrt{\frac{\kappa}{1+\kappa}}\mathbf H_{\mathrm{LoS}}
+\sqrt{\frac{1}{1+\kappa}}\mathbf H_{\mathrm{NLoS}},
\label{eq:rician}
\end{equation}
where $\kappa$ is the Rician K-factor, $\mathbf H_{\mathrm{LoS}}$ is the line-of-sight (LoS) component, and $\mathbf H_{\mathrm{NLoS}}$  is the non line-of-sight (NLoS) component of the channel 
\footnote{The proposed beamforming framework can be applied to any MIMO communication channel model.}.

The LoS component is written as \cite{alkhateeb} 
\begin{equation}
\mathbf H_{\mathrm{LoS}}=\sqrt{N_\mathrm{T}N_\mathrm{R}}\,
\mathbf a_\mathrm{R}(\phi)\mathbf a_\mathrm{T}^H(\phi),
\label{eq:los}
\end{equation}
where $\phi$ is the DoD and DoA at Alice and Bob, respectively, for the LoS component. The NLoS component is written as 
\begin{equation}
\mathbf H_{\mathrm{NLoS}}=
\sqrt{\frac{N_\mathrm{T}N_\mathrm{R}}{L_p}}
\sum_{\ell=1}^{L_p}\alpha_\ell
\mathbf a_\mathrm{R}(\omega_{r,\ell})\mathbf a_\mathrm{T}^H(\omega_{t,\ell}),
\label{eq:nlos}
\end{equation}
where $L_p$ denotes the number of NLoS paths, $\alpha_\ell\sim\mathcal{CN}(0,1)$ is the path gain for each NLoS path $\ell$, and $\omega_{t,\ell} \in [-90,90]^\circ$ and $\omega_{r,\ell}\in [-90,90]^\circ$ are the DoD and DoA at Alice and Bob, respectively, for each NLoS path $\ell$.



The achievable rate for the communication channel is written as 
\begin{equation}
C(\mathbf Q) =\log_2\det\!\left(
\mathbf I_{N_\mathrm{R}}+\sigma_\mathrm{c}^{-2}\mathbf H_\mathrm{c}\mathbf Q\mathbf H_\mathrm{c}^H
\right)\quad\text{bps/Hz}.
\label{eq:rate}
\end{equation}

\subsection{Sensing Model}
We consider MIMO radar sensing over a particular coherent processing interval (CPI) containing $L>N_\mathrm{T}$ symbols. The transmitted signal matrix in this interval is written as
\begin{equation}
\mathbf X=[\mathbf x(1),\ldots,\mathbf x(L)]\in\mathbb C^{N_\mathrm{T}\times L}.
\label{eq:block}
\end{equation}
For sufficiently large $L$, $L^{-1}\mathbf X\mathbf X^H\approx\mathbf Q$~\cite{liuCRB}. The sensing echo matrix at Alice for a single far-field point target can be written as 
\begin{equation}
\mathbf Y_\textrm{s}=\mathbf H_\textrm{s}\mathbf X+\mathbf Z_\textrm{s},
\label{eq:echo}
\end{equation}
where $\mathbf H_\textrm{s}$ denotes the sensing channel and $\mathbf Z_\textrm{s}\in\mathbb C^{N_\textrm{S}\times L}$ denotes the sensing-receiver noise matrix, whose entries are independent circularly symmetric complex Gaussian random variables with mean zero and  variance $\sigma_\textrm{s}^2$. Using the same antenna steering vector model as in the communication model in section \ref{sec_comm_model}, the sensing channel is written as \cite{liuCRB}
\begin{equation}
\mathbf H_\textrm{s}=\alpha\mathbf A(\theta)=\alpha\mathbf b(\theta)\mathbf a_\mathrm{T}^H(\theta),
\label{eq:target}
\end{equation} 
where $\theta$ is the target angle relative to Alice's antenna array,  $\alpha\in\mathbb C$ is the reflection coefficient, which depends on the round-trip path loss between Alice and the target and the radar cross section (RCS) of the target \cite{crbrate}, $\mathbf b(\theta)$ is the sensing array
steering vector defined as in \eqref{eq_steering_vector} with $N_\textrm{S}$ elements, and  $\ {\mathbf A}(\theta)=\mathbf b(\theta)\mathbf a_\mathrm{T}^H(\theta)$ is the target response matrix. 


The angle-estimation Cram\'er--Rao bound (CRB) \footnote{Here we omit the angle argument $\theta$ for brevity.} is \cite{liuCRB}
\begin{equation}
\begin{aligned}
&\mathrm{CRB(\mathbf Q)}\\
&=\frac{\sigma_\textrm{s}^2}{2|\alpha|^2L}
\frac{\operatorname{tr}(\mathbf A^H\mathbf A\mathbf Q)}
{\operatorname{tr}(\dot{\mathbf A}^H\dot{\mathbf A}\mathbf Q)
\operatorname{tr}(\mathbf A^H\mathbf A\mathbf Q)
-|\operatorname{tr}(\dot{\mathbf A}^H\mathbf A\mathbf Q)|^2},
\end{aligned}
\label{eq:crb}
\end{equation}
where $\dot{\mathbf A}(\theta)$ denotes the derivative of
$\mathbf A(\theta)$ with respect to $\theta$ in radians, given by
$\dot{\mathbf A}(\theta)
=\dot{\mathbf b}(\theta)\mathbf a_\mathrm{T}^H(\theta)
+\mathbf b(\theta)\dot{\mathbf a}_\mathrm{T}^H(\theta)$.
Here, $\dot{\mathbf a}_\mathrm{T}(\theta)$ and
$\dot{\mathbf b}(\theta)$ denote the derivatives of
$\mathbf a_\mathrm{T}(\theta)$ and $\mathbf b(\theta)$,
respectively.



\color{black}


To evaluate the CRB, we use coarse estimates of $\theta$ and $|\alpha|$ obtained in a preceding sensing stage. This corresponds to target tracking, where a static or slowly moving target permits these estimates to remain sufficiently accurate for the current waveform design~\cite{liuCRB}.

\subsection{Transmitter location privacy model}
To preserve Alice's location privacy from Bob, we use the direction-of-arrival obfuscation power ratio (DoA-OPR) introduced in \cite{privacy_our_paper}. This metric is defined as the ratio of the received angular power in a prescribed fake direction $\hat\phi$ to that in the actual direction $\phi$, expressed as  
\begin{align} 
\gamma(\mathbf Q) &=
\frac{\mathbf a_\mathrm{R}^H(\hat\phi)\mathbf R \mathbf a_\mathrm{R}(\hat\phi)}
{\mathbf a_\mathrm{R}^H(\phi)\mathbf R \mathbf a_\mathrm{R}(\phi)},
\label{eq:privacy}
\end{align}
where the covariance matrix of the received signal at Bob is
\begin{align} 
\mathbf R &=\mathbf H_\textrm{c}\mathbf Q\mathbf H_\textrm{c}^H+\sigma_\textrm{c}^2\mathbf I_{N_\mathrm{R}}.
\label{eq:received_cov}
\end{align}
  Thus, $\gamma >1$ means that the fake direction has a greater received angular power than the actual direction. The DoA-OPR enables Alice to shape the received power at Bob to achieve location privacy. 
The DoA-OPR can be alternatively expressed as \cite{privacy_our_paper}
\begin{align}
\label{er_gamma_rayleighQ_form}
\gamma(\mathbf Q) 
=\frac{\text{tr}(\mathbf{W}^H \mathbf{A}_{\text{fake}} \mathbf{W})}{\text{tr}(\mathbf{W}^H \mathbf{A}_{\text{actual}} \mathbf{W})},
\end{align}
where 
\begin{align}
\mathbf A_{\mathrm{fake}}&=
\mathbf H_\textrm{c}^H\mathbf a_\mathrm{R}(\hat\phi)\mathbf a_\mathrm{R}^H(\hat\phi)\mathbf H_\textrm{c}
+\frac{\sigma_\textrm{c}^2}{P}\mathbf I_{N_\mathrm{T}},\label{eq:Af}\\
\mathbf A_{\mathrm{actual}}&=
\mathbf H_\textrm{c}^H\mathbf a_\mathrm{R}(\phi)\mathbf a_\mathrm{R}^H(\phi)\mathbf H_\textrm{c}
+\frac{\sigma_\textrm{c}^2}{P}\mathbf I_{N_\mathrm{T}}.\label{eq:Aa}
\end{align}



 

\section{Problem Formulation}
\label{sec:formulation}
To design the precoder $\mathbf{W}$ at Alice, or equivalently the transmit covariance $\mathbf Q$, we formulate an achievable rate maximization problem subject to the transmit power, privacy, and sensing requirements. Specifically, the optimization problem is written as
\begin{subequations}\label{eq:P1}
\begin{align}
\text{P1:}\quad\max_{\mathbf Q\succeq\mathbf0}\quad
&C(\mathbf Q)\\
\text{s.t.}\quad
&\operatorname{tr}(\mathbf Q)=P,
\label{eq:power_constraint}\\
&\gamma(\mathbf Q)\ge\gamma_{\mathrm{pri}},
\label{eq:privacy_constraint}\\
&\mathrm{CRB}(\mathbf Q)\le\gamma_{\mathrm{CRB}},
\label{eq:sensing_constraint}\\
&\operatorname{rank}(\mathbf Q)=N_{\mathrm d},
\label{eq:rank_constraint}
\end{align}
\end{subequations}
where~\eqref{eq:power_constraint} enforces the total transmit
power $P$, \eqref{eq:privacy_constraint} ensures the desired
privacy threshold $\gamma_{\mathrm{pri}},
$ \eqref{eq:sensing_constraint} imposes the CRB
threshold $\gamma_{\mathrm{CRB}}$, and
\eqref{eq:rank_constraint} specifies the number of active
data streams $N_{\mathrm d}$. The privacy constraint~\eqref{eq:privacy_constraint}
can be equivalently expressed as \cite{privacy_our_paper}
\begin{equation}
\operatorname{tr}\!\left[
(\mathbf A_{\mathrm{fake}}
-\gamma_{\mathrm{pri}}\mathbf A_{\mathrm{actual}})
\mathbf Q\right]\ge0.
\label{eq:privacy_linear}
\end{equation}
Although $C(\mathbf{Q})$ is concave in $\mathbf Q$,
P1 is nonconvex due to the rank constraint in \eqref{eq:rank_constraint}.

\color{black}

\section{ Rate--Sensing--Privacy Tradeoff }
\label{sec:region}
Our objective in this section is to analyze the three-way tradeoff among communication rate, target-sensing performance, and transmitter location privacy towards the solution of P1. This is accomplished by first finding the feasible values of the privacy threshold, and then, for each privacy threshold, finding the tradeoff between the CRB and the communication rate performance.

\subsection{Feasibility of privacy threshold $\gamma_{\mathrm{pri}}$}

The feasibility of the optimization problem depends on $\gamma_{\textrm{pri}}$ in P1. Thus, we find the maximum and minimum possible values of $\gamma$, respectively. 
As $\mathbf{A}_{\text{fake}}\in \mathbb{C}^{N_\textrm{T} \times N_\textrm{T}}$ and $\mathbf{A}_{\text{actual}}\in \mathbb{C}^{N_\textrm{T} \times N_\textrm{T}}$ are positive definite and Hermitian, $\gamma$ in (\ref{er_gamma_rayleighQ_form}) is in the form of a generalized Rayleigh quotient. Thus, the feasible range of $\gamma$ is defined as

\begin{align}
\label{eq_limits_on_gamma}
\gamma_{\min} \leq \gamma\leq \gamma_{\max},
\end{align}
where $\gamma_{\min}$  and $\gamma_{\max}$ denote the smallest and largest generalized eigenvalue of the matrix pair \{$\mathbf{A}_{\text{fake}}$, $\mathbf{A}_{\text{actual}}$\}, respectively {\cite{Channel_Estimation}.
The privacy constraint is
infeasible when $\gamma_{\mathrm{pri}}>\gamma_{\max}$ and
inactive when $\gamma_{\mathrm{pri}}<\gamma_{\min}$.





\subsection{  CRB-Rate region characterization  }
For a given feasible privacy threshold, we characterize the tradeoff between sensing CRB  and communication rate under the transmit-power and data-stream requirements. To characterize the CRB-rate region for a given privacy threshold, we first obtain two corner points on each boundary by minimizing CRB for sensing only and maximizing rate for communication only, respectively.

The CRB minimization corner point is obtained by minimizing the sensing
CRB under the power, privacy, and rank constraints of P1, i.e.,  
\begin{align}
\text{P2:}\quad
\underset{\mathbf Q\succeq\mathbf0}{\min}\quad
&\mathrm{CRB}(\mathbf Q)
\label{eq:sensing_corner}\\
\text{s.t.}\quad
&\operatorname{tr}(\mathbf Q)=P,\;
\gamma(\mathbf Q)\ge\gamma_{\mathrm{pri}},\;
\operatorname{rank}(\mathbf Q)=N_\mathrm{d}.\nonumber
\end{align}
The resulting transmit covariance $\mathbf Q_s^*$ gives the sensing only
corner $(\zeta_\mathrm{s},C_\mathrm{s})$, where
\begin{equation}
\zeta_\mathrm{s}=\mathrm{CRB}(\mathbf Q_\mathrm{s}^*),
\qquad C_\mathrm{s}=C(\mathbf Q_\mathrm{s}^*).
\label{eq:sensing_corner_pair}
\end{equation}

Under the same constraints, the rate maximization corner point is obtained
by maximizing the achievable rate,
\begin{align}
\text{P3:}\quad
\underset{\mathbf Q\succeq\mathbf0}{\max}\quad
&C(\mathbf Q)
\label{eq:comm_corner}\\
\text{s.t.}\quad
&\operatorname{tr}(\mathbf Q)=P,\;
\gamma(\mathbf Q)\ge\gamma_{\mathrm{pri}},\;
\operatorname{rank}(\mathbf Q)=N_\mathrm{d}.\nonumber
\end{align}
The resulting transmit covariance $\mathbf Q_\mathrm{c}^*$ gives the communication only
corner $(\zeta_\mathrm{c},C_\mathrm{c})$, where
\begin{equation}
\zeta_\mathrm{c}=\mathrm{CRB}(\mathbf Q_\mathrm{c}^*),
\qquad C_\mathrm{c}=C(\mathbf Q_\mathrm{c}^*).
\label{eq:comm_corner_pair}
\end{equation}
Both problems are nonconvex due to the rank constraint and the solutions are obtained via the proposed methodology in Section \ref{sec_solution}. 
Finally, to find the boundary points between the two corners on the CRB-rate region, we vary the CRB threshold from \(\zeta_{\mathrm{s}}\) to \(\zeta_{\mathrm{c}}\) and solve P1 for each threshold using the proposed method.

\section{Solution Strategy}
\label{sec_solution}

In this section, we first reformulate the fractional CRB constraint in~\eqref{eq:sensing_constraint} as an equivalent linear matrix inequality (LMI) to transform it into
standard semidefinite form. Then, to address the nonconvex rank constraint \eqref{eq:rank_constraint} in P1, we select exactly $N_d$ eigenmodes and optimize their power allocations.






The CRB constraint in~\eqref{eq:sensing_constraint} is a fractional function of the transmit covariance. To obtain an equivalent LMI form with entries affine in \(\mathbf Q\), we substitute~\eqref{eq:crb} into~\eqref{eq:sensing_constraint} and apply the Schur complement~\cite{boyd}, yielding 
\begin{equation}
\begin{bmatrix}
\operatorname{tr}(\dot{\mathbf A}^H\dot{\mathbf A}\mathbf Q)-\widetilde\gamma_\mathrm{CRB}^{-1}
&\operatorname{tr}(\dot{\mathbf A}^H\mathbf A\mathbf Q)\\
\operatorname{tr}(\mathbf A^H\dot{\mathbf A}\mathbf Q)
&\operatorname{tr}(\mathbf A^H\mathbf A\mathbf Q)
\end{bmatrix}\succeq\mathbf0,
\label{eq:lmi}
\end{equation}
where, for notational convenience, we define
$\widetilde{\gamma}_{\mathrm{CRB}}
\triangleq 2\gamma_{\mathrm{CRB}}L|\alpha|^2/\sigma_{\mathrm{s}}^2$.
The resulting LMI is equivalent to the sensing constraint in
\eqref{eq:sensing_constraint}.


To handle the nonconvex rank constraint, we begin with the
eigendecomposition of the Hermitian matrix
$\mathbf A_{\mathrm{fake}}-\gamma_{\mathrm{pri}}\mathbf A_{\mathrm{actual}}$
in~\eqref{eq:privacy_linear}:

\begin{equation}
\label{eq_eigen_modes}
\mathbf A_{\mathrm{fake}}-\gamma_{\mathrm{pri}}\mathbf A_{\mathrm{actual}}=\mathbf U\boldsymbol\Lambda\mathbf U^H,
\end{equation}
where $\mathbf U$ denotes the unitary eigenvector matrix and $\boldsymbol\Lambda=\operatorname{diag}(\lambda_1,\ldots,\lambda_{N_\mathrm{T}})$ is the diagonal matrix of real eigenvalues. 
Thus, to simplify the solution, we choose the Hermitian covariance $\mathbf Q$ to commute with $(\mathbf A_{\mathrm{fake}}-\gamma_{\mathrm{pri}}\mathbf A_{\mathrm{actual}})$ \cite{privacy_our_paper}. This design choice makes the two matrices simultaneously diagonalizable and share the same unitary
matrix $\mathbf U$. For a chosen common unitary eigenbasis $\mathbf U$, we assume $\mathbf P$ as a diagonal power-allocation matrix such that
\begin{equation}
\mathbf Q=\mathbf U\mathbf P\mathbf U^H
\label{eq:structured}
\end{equation}
where
$\mathbf{P} = \text{diag}(p_1, p_2, \ldots, p_{N_\textrm{T}})$ is a diagonal matrix with real non-negative entries, $\text{tr}
(\mathbf{P})  =  P$, and $\textrm{rank} (\mathbf{P}) = N_\textrm{d}$. Due to the rank constraint on $\mathbf{Q}$, out of $N_\textrm{T}$ diagonal entries of $\mathbf{P}$, only $N_\mathrm{d}$ will be non-zero.




We modify P1 under the chosen covariance in~\eqref{eq:structured} and use \eqref{eq:lmi} and \eqref{eq_eigen_modes} to obtain
\begin{subequations}
\label{eq:P2}
\begin{align}
\text{P4:}  &\max_{\small{\begin{array}{l}
            \mathcal{I}, \{ p_i \}_{i \in \mathcal{I}}
        \end{array}}} 
\log_2\det\!\left(\mathbf I_{N_\mathrm{R}}+
\frac{\mathbf H_\mathrm{c}\mathbf U_{\mathcal I}\mathbf P_{\mathcal I}\mathbf U_{\mathcal I}^H\mathbf H_\mathrm{c}^H}{\sigma_\mathrm{c}^{2}}\right)
\label{eq:P2obj}\\
\text{s.t.}&\quad \sum p_i=P, \quad p_i>0, \label{eq:P2power}\\
&\quad\sum \lambda_i p_i\ge0,\label{eq:P2privacy}\\
&\quad\begin{bmatrix}
\displaystyle\sum p_i b_{DD,i}-\widetilde\gamma_\mathrm{CRB}^{-1}
&\quad\displaystyle\sum p_i b_{DA,i}\\
\displaystyle\overline{\left(\sum p_i b_{DA,i}\right)}
&\quad\displaystyle\sum p_i b_{AA,i}
\end{bmatrix}\succeq\mathbf0,
\label{eq:P2crb}\\
&\quad\mathcal I\subseteq\{1,\ldots,N_\mathrm{T}\},\quad |\mathcal I|=N_\mathrm{d},\label{eq:P2subset}\\
&\quad \forall i\in\mathcal I.
\label{eq:P2positive}
\end{align}
\end{subequations}
where $\mathcal{I}$ denotes the set (of size $N_\textrm{d}$) indices of the positive diagonal entries of $\mathbf{P}$, and  $\mathbf{U}_{\mathcal{I}}$ and $\mathbf{P}_{\mathcal{I}}$ are the submatrices taking columns with indices in $\mathcal{I}$ from $\mathbf{U}$ and $\mathbf{P}$, respectively. The elements of $\mathbf{P}_{\mathcal{I}}$ are the allocated powers to the $N_\textrm{d}$ eigenvectors in (\ref{eq_eigen_modes}). 
The constraint in (\ref{eq:P2privacy}) is obtained from (\ref{eq:privacy_linear}) with the help of (\ref{eq_eigen_modes}) and (\ref{eq:structured}),  and (\ref{eq:P2crb}) is obtained from (\ref{eq:lmi}) 
using the coefficients defined as 
\begin{align}
b_{DD,i} &=\mathbf u_i^H\dot{\mathbf A}^H\dot{\mathbf A}\mathbf u_i, 
b_{DA,i} =\mathbf u_i^H\dot{\mathbf A}^H\mathbf A\mathbf u_i, \nn\\
b_{AA,i} &=\mathbf u_i^H\mathbf A^H\mathbf A\mathbf u_i.
\label{eq:projected}
\end{align}
where $\mathbf u_i$ denotes the $i$th column of $\mathbf U$. 

The solution of P4 boils down to finding jointly 
the best set of indices $\mathcal{I}^*$ of  $N_\mathrm{d}$ eigenvectors in (\ref{eq_eigen_modes}) and the optimal
 allocation of power $\mathbf{P}_{\mathcal{I}^*}$ to the combination of eigenvectors so
that the achievable rate in (\ref{eq:P2obj}) is maximized. As the achievable rate is concave in $  \mathbf{P}_{\mathcal{I}}$ and the power, privacy, and sensing constraints are convex for the subset $\mathcal I$, a convex optimization procedure would provide the optimal power allocation for that  subset.
Then an exhaustive search on $\binom{N_\mathrm{T}}{N_\mathrm{d}}$ eigenmode subsets would be required to find the best subset $\mathcal I^*$ with the highest rate. The resulting transmit covariance and precoder are given by
\begin{equation}
\mathbf Q^*=\mathbf U_{\mathcal I^*}\mathbf P_{\mathcal I^*}
\mathbf U_{\mathcal I^*}^{H},
\qquad
\mathbf W^*=\mathbf U_{\mathcal I^*}\mathbf P_{\mathcal I^*}^{1/2}.
\label{eq:Qstar_Wstar}
\end{equation}




\section{Eigenmode Subset Pruning}
This section discusses the way to reduce the computational cost towards achieving the solution of P4. The problem P4 requires invoking a convex optimization
procedure for each of $\binom{N_\mathrm {T}}{N_\mathrm{d}}$ eigenmode subsets.
To reduce this computational cost, we apply two eigenmode subset pruning strategies to
discard infeasible subsets before invoking a convex optimization
procedure.

\textit{Privacy-based pruning:}
A subset $\mathcal I$ is discarded if $\lambda_i<0$ for every
$i\in\mathcal I$, since the allocated powers are positive, such a subset violates the constraint
$\sum_{i\in\mathcal I}\lambda_i p_i\ge0$ 
in \eqref{eq:P2privacy}.


\textit{CRB-based pruning:}
Since the matrix in \eqref{eq:P2crb} must be positive semidefinite, it should have nonnegative diagonal entries.
Therefore, the first diagonal entry of~\eqref{eq:P2crb} gives
the necessary condition for feasibility
\begin{equation}
\sum_{i\in\mathcal I}p_i b_{DD,i}
\ge\widetilde{\gamma}_\mathrm{CRB}^{-1}.
\label{eq:filter_necessary}
\end{equation}
From the power constraints in \eqref{eq:P2power}, we must have

\begin{equation}
\sum_{i\in\mathcal I}p_i b_{DD,i}
\le
\left(\max_{i\in\mathcal I}b_{DD,i}\right)
\sum_{i\in\mathcal I}p_i
=
P\max_{i\in\mathcal I}b_{DD,i}.
\label{eq:filter_bound}
\end{equation}
Combining~\eqref{eq:filter_necessary}
and~\eqref{eq:filter_bound}, a subset is discarded if
\begin{equation}
P\max_{i\in\mathcal I}b_{DD,i}
<\widetilde{\gamma}_\mathrm{CRB}^{-1}.
\label{eq:filter_reject}
\end{equation}
After pruning, P4 is solved for each remaining eigenmode subset.

\section{{Sensing-Aligned} Fake Angle Selection}
\label{SAFA}
Jointly satisfying sensing and privacy requirements leaves less power for communication, which incurs a rate penalty. To mitigate this rate penalty, we propose a fake angle selection technique that exploits power directed
toward the target to also contribute to the power in the fake direction.
For a fixed sensing-target angle $\theta$, we identify
the NLoS path whose angle of departure is closest to $\theta$
and assign its corresponding angle of arrival at Bob as the
fake direction\footnote{Here, Alice is assumed to know the angular structure of
the channel. This assumption is not necessary for the solution strategy in section \ref{sec_solution}.}. 
Precisely, the NLoS path and fake angle selected, respectively, are
\begin{align}
\label{eq_fake_angle}
\ell^*
=\arg\min_{\ell\in\{1,\ldots,L_p\}}
\left|\omega_{t,\ell}-\theta\right|,
~~\textrm{and}~~
\hat{\phi}=\omega_{r,\ell^*}.
\end{align}
Transmit power for target sensing
aids location privacy through the NLoS path selected in (\ref{eq_fake_angle}), leaving more power for communication. This reduces the communication rate penalty of jointly satisfying the sensing and privacy constraints. We refer to this as the sensing-aligned fake angle selection technique.

\section{Results and Discussion}

In this section, we present numerical results. Unless otherwise specified,
the parameter values considered for the numerical results are $N_{\mathrm T}=N_{\mathrm R}=8$, $N_{\mathrm S}=12$,
$N_\mathrm{d}=6$, $L_p=30$, $L=200$, $|\alpha|=5\times10^{-3}$, $\sigma_\mathrm {c}^2=\sigma_\mathrm {s}^2=1$, $\phi=20^\circ$, $\hat{\phi}=50^\circ$, and $\theta=-45^\circ$. Figs.~2 and~5 show averaged results, whereas Figs.~3 and~4
use distinct single channel realization.



\subsection{Rate--Sensing--Privacy Tradeoff}

Fig. \ref{fig:rate_crb_privacy_3d} presents a three-dimensional
view of the joint dependence of achievable rate on sensing
and privacy requirements, averaged over 100 channel realizations.
The achievable rate decreases as either requirement becomes
more stringent, since satisfying these requirements restricts
the transmit covariance under the shared power budget.
This behavior illustrates the three-way tradeoff among
communication rate, target-sensing accuracy, and transmitter
location privacy.

Fig. \ref{fig:rate_crb_2d} presents the corresponding
two-dimensional view for $\gamma_{\mathrm{pri}}\in\{0,1,2\}$.
For each privacy threshold, relaxing the sensing requirement
increases the rate from the sensing corner
$(\zeta_{\mathrm{s}},C_{\mathrm{s}})$ toward the
communication corner $(\zeta_{\mathrm{c}},C_{\mathrm{c}})$,
beyond which further relaxation provides no additional rate gain.
At $\gamma_\mathrm{CRB}=6\times10^{-3}\,\mathrm{rad}^{2}$, the average rates are
$34.2$, $28.5$, and $28.0$ bits/s/Hz (bps/Hz) for
$\gamma_{\mathrm{pri}}=0,1,2$, respectively, quantifying the
communication penalty of enforcing stronger privacy at the
same sensing accuracy.

\begin{figure}[t]
    \centering
    \begin{subfigure}[b]{0.48\linewidth}
        \centering
        \includegraphics[width=\linewidth]{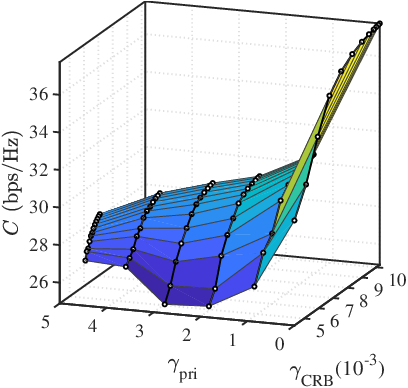}
        \caption{3D tradeoff surface}
        \label{fig:rate_crb_privacy_3d}
    \end{subfigure}
    \hfill
    \begin{subfigure}[b]{0.48\linewidth}
        \centering
        \includegraphics[width=\linewidth]{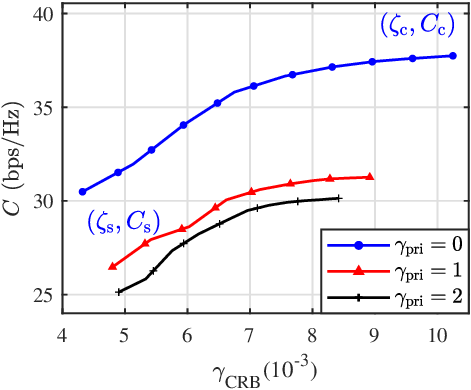}
        \caption{2D Rate--CRB tradeoff}
        \label{fig:rate_crb_2d}
    \end{subfigure}
    \caption{Rate--sensing--privacy tradeoff.}
    \label{fig:rate_crb_privacy}
\end{figure}

\begin{figure}[t]
    \centering
    \begin{subfigure}[t]{0.48\linewidth}
        \centering
        \includegraphics[width=\linewidth]{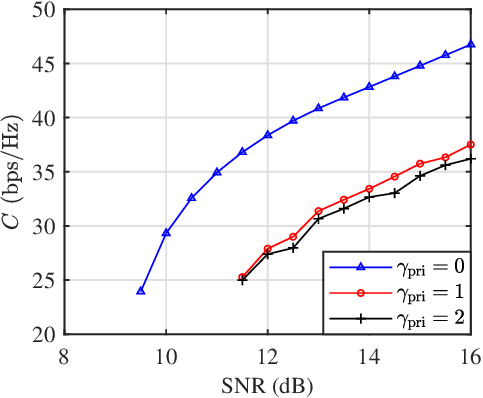}
        \caption{Achievable rate versus SNR.}
        \label{fig:privacy_rate}
    \end{subfigure}
    \hfill
    \begin{subfigure}[t]{0.48\linewidth}
        \centering
        \includegraphics[width=\linewidth]{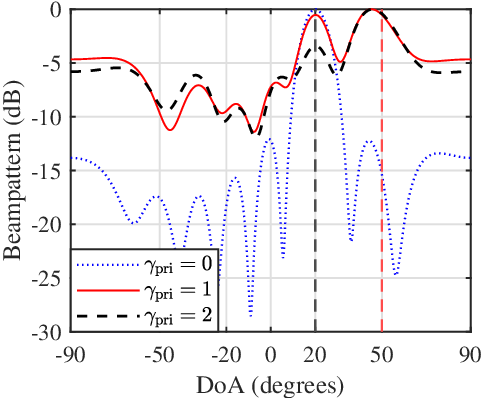}
        \caption{Beampattern at SNR = 13~dB.}
        \label{fig:privacy_beampattern}
    \end{subfigure}
    \caption{Impact of the privacy threshold with $\gamma_{\mathrm{CRB}}=4\times10^{-3}\,\mathrm{rad}^{2}$.}
    \label{fig:privacy_threshold}
\end{figure}

\begin{figure}[t]
    \centering
    \begin{subfigure}[t]{0.48\linewidth}
        \centering        \includegraphics[width=\linewidth]{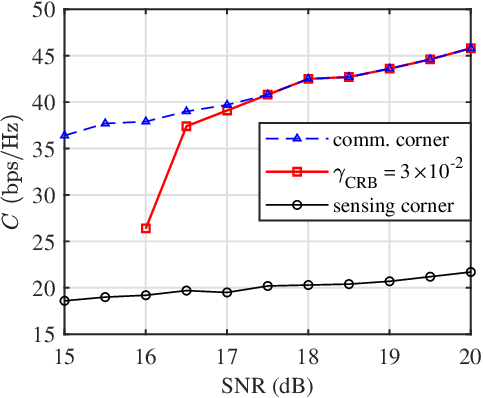}
        \caption{Achievable rate versus SNR.}
        \label{fig:crb_rate}
    \end{subfigure}
    \hfill
    \begin{subfigure}[t]{0.48\linewidth}
        \centering        \includegraphics[width=\linewidth]{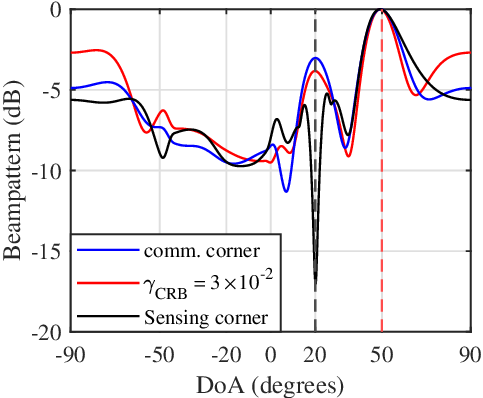}
        \caption{Beampattern at SNR = 16~dB.}
        \label{fig:crb_beampattern}
    \end{subfigure}
    \caption{Impact of the CRB requirement with $\gamma_{\mathrm{pri}}=2$.}
    \label{fig:crb_cases}  
    \vspace{-.25cm}
\end{figure}
\subsection{Impact of the Privacy Threshold}

Figs. \ref{fig:privacy_rate}   and \ref{fig:privacy_beampattern} illustrate the achievable rate versus signal-to-noise ratio (SNR) $= P/\sigma_\mathrm{c}^2$ and the received beampattern at Bob, respectively, for a single channel realization with different privacy thresholds under a fixed CRB requirement. Fig. \ref{fig:privacy_rate} shows that the achievable rate increases with SNR in all cases. A minimum SNR is required to jointly satisfy the sensing and privacy constraints in problem P4, ensuring feasible operation. As the privacy threshold increases from $\gamma_{\mathrm{pri}}=0$ to $1$, the minimum SNR required for feasible operation increases from approximately 9.5 to 11.5~dB. This increase reflects the additional power required to enforce privacy while satisfying the same sensing requirement.

In  Fig. \ref{fig:privacy_beampattern}, $\gamma_{\mathrm{pri}}=0$ (no privacy constraint)
yields a dominant peak in the actual direction.
For $\gamma_{\mathrm{pri}}=1$, the powers in the actual and
fake directions become comparable, whereas
$\gamma_{\mathrm{pri}}=2$ makes the fake direction dominant.
This angular power distribution strengthens location privacy
but at the expense of achievable rate, as reflected in Fig. \ref{fig:privacy_rate}.


\subsection{Impact of the Sensing Requirement}
Fig. \ref{fig:crb_rate} plots the achievable rate versus SNR for different
CRB requirements when $\gamma_{\mathrm{pri}}=2$

The sensing only and
communication only designs provide the lower and upper rate bounds, respectively. For
$\gamma_{\mathrm{CRB}}=3\times10^{-2}\,\mathrm{rad}^{2}$,
feasible operation begins at approximately 16~dB.
At lower SNRs, the prescribed CRB threshold is below the
minimum achievable CRB under the imposed privacy requirement,
making the problem infeasible. As SNR increases, the
CRB requirement can be easily met, and the achievable
rate approaches the rate upper bound.

Fig. \ref{fig:crb_beampattern} shows the received beampattern at Bob corresponding to Fig. \ref{fig:crb_rate}.
All three designs provide greater power in the fake direction relative to the
actual direction as $\gamma_{\mathrm{pri}}=2$. 
Moving from the communication corner to the sensing corner, the power in the actual direction decreases as the sensing requirement becomes progressively more stringent. Excessive power suppression in the actual direction may reveal the transmitter's location to Bob. Hence, higher sensing requirements may compromise the transmitter’s location privacy.






\subsection{Sensing-Aligned Fake-Angle Selection}

In Fig. \ref{fig:rate_privacy_safa}, we compare the average achievable rate of the proposed sensing-aligned 
fake-angle selection technique with random fake-angle selection by varying privacy thresholds for different CRB thresholds $5\times 10^{-3}$ and
$8\times 10^{-3}\,\mathrm{rad}^{2}$. 
The proposed technique achieves a higher rate than the random fake angle selection technique except at $\gamma_{\mathrm{pri}}=0$.
Both techniques achieve the same rate at $\gamma_{\mathrm{pri}} = 0$ as the privacy constraint is inactive. 
The achievable rate decreases with increasing privacy threshold. However, as the privacy threshold increases, the rate gain of the proposed  technique over the random fake angle selection technique increases.
As the sensing requirement becomes more stringent, i.e., $\gamma_{\mathrm{CRB}}$  decreases from $8\times10^{-3}$ to
$5\times10^{-3}\,\mathrm{rad}^{2}$, the rate difference between the two techniques increases as well.
These observations indicate that the proposed technique is particularly
beneficial when sensing and privacy requirements are stringent.



\begin{figure}
    \centering    \includegraphics[width=0.5\linewidth]{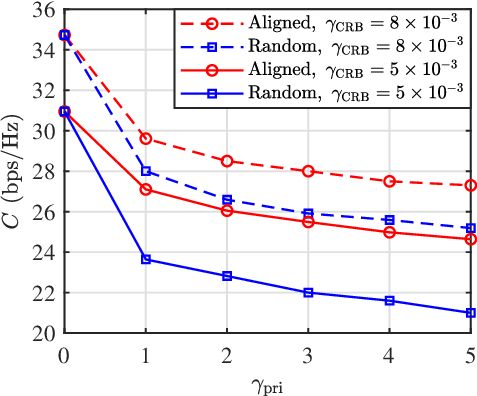}
    \caption{Average rate versus privacy threshold for the proposed and random fake angle selection techniques.}
    \label{fig:rate_privacy_safa}
    \vspace{-.5cm}
\end{figure}

\section{Conclusion} In this paper, a transmit beamformer is designed to preserve the location privacy of a MIMO ISAC transmitter communicating with a receiver while sensing a point target. An achievable rate maximization problem is formulated 
which is nonconvex and the solution is provided  through an eigenmode-selection and power-allocation method. The computational complexity of the solution is reduced by discarding
infeasible candidate solutions. 
Furthermore, a sensing-aligned fake angle selection technique is proposed that exploits the same spatial power for both target sensing and location privacy. Numerical results show that more stringent sensing or privacy requirements reduce the achievable rate.
The proposed
fake angle selection technique improves the achievable rate and is more
beneficial when sensing and privacy requirements are stringent.





\section*{Acknowledgment}
This work was supported in part by Taighde Éireann – Research Ireland under Grant numbers 22/PATH-S/10788 and 24/FFP-P/12895.
\vspace{-.2cm}
\bibliographystyle{IEEEtran}
\bibliography{references}

\end{document}